\documentclass[aps,11pt,a4paper]{article}
\usepackage{amsfonts}
\usepackage{amsmath}
\usepackage{amssymb}
\usepackage{upgreek}
\usepackage{bm}
\usepackage[colorlinks=true,linkcolor=blue]{hyperref}
\usepackage{url}
\usepackage{txfonts}
\usepackage[T1]{fontenc}
\usepackage{graphicx}
\usepackage{authblk}

\newcommand\rd{\text{d}}
\newcommand\re{\text{e}}

\newcommand\rrp{\text{p}}
\newcommand\rH{\text{H}}
\newcommand\rn{\text{n}}
\newcommand\rD{\text{D}}
\newcommand\rT{\text{T}}

\newcommand\rhe{{}^3\text{He}}

\newcommand\ra{{}^4\text{He}}
\newcommand\rli{{}^6\text{Li}}
\newcommand\rLI{{}^7\text{Li}}
\newcommand\rBE{{}^7\text{Be}}
\newcommand\libe{\rLI/\rBE}
\newcommand\rB{{}^{10}\text{B}}

\begin{document}
\title{Non-Thermal Cosmic Rays During Big Bang Nucleosynthesis to Solve the Lithium Problem}

\author[a,b]{Ming-Ming Kang}
\affil[a]{College of Physical Science and Technology, Sichuan University, Chengdu 610064, P.R. China}
\affil[b]{Key Laboratory of Radiation Physics and Technology of Ministry of Education, Sichuan University, Chengdu 610064, P.R. China}
\author[c]{Yang Hu\thanks{Corresponding author: yanghu@shmtu.edu.cn}}
\affil[c]{College of Arts and Sciences, Shanghai Maritime University, Shanghai 201306, P.R. China}
\author[d,e]{Hong-Bo Hu}
\affil[d]{Key Laboratory of Particle Astrophysics, Institute of High Energy Physics, Chinese Academy of Sciences, Beijing 100049, P.R. China}
\affil[e]{University of Chinese Academy of Sciences, Beijing 100049, P.R. China}
\author[f,g,h]{and Shou-Hua Zhu}
\affil[f]{Institute of Theoretical Physics $\&$ State Key Laboratory of Nuclear Physics and Technology, Peking University, Beijing 100871, P.R. China}
\affil[g]{Collaborative Innovation Center of Quantum Matter, Beijing 100871, P.R. China}
\affil[h]{Center for High Energy Physics, Peking University, Beijing 100871, P.R. China}

\date{}
\maketitle

\begin{abstract}
The discrepancy between the theoretical prediction of primordial lithium abundances and astronomical observations is called the Lithium Problem.
We find that extra contributions from non-thermal hydrogen and helium during Big Bang nucleosynthesis can explain the discrepancy,
for both Li-7 and Li-6,
and will change the deuterium abundance only little.
The allowed parameter space of such an amount of non-thermal particles and the energy range is shown.
The hypothesis is stable regardless of the cross-section uncertainty of relevant reactions and the explicit shape of the energy spectrum.
\end{abstract}

\section{Introduction}
Standard Big Bang nucleosynthesis (SBBN) is one of the three pillars of the Big Bang cosmology\cite{1505.01076}.
As a powerful tool with which to study the early Universe, the primordial light element abundances,
such as D/H and He-4 ($\alpha$ particle), are in close agreement with astronomical observations, except for lithium.
Specifically, the theoretically predicted Li-7 abundance is about 3 times higher
than indicated by measurements from metal-poor galactic halo stars,
while the Li-6 abundance is about 2 orders of magnitude lower than observations.
The discrepancy, which is called the Lithium Problem, is still unexplained\cite{9611043,9905211,Cyburt}.
(Also see reviews in \cite{Iocco,1011.1054,1203.3551}.)

Extensive investigations have been undertaken in attempts to solve the Lithium Problem,
from the points of view of astronomical and astrophyical origin, nuclear reactions, cosmological parameters, modifications of the SBBN, etc.
Nuclear processes in stars will change the lithium abundances,
so primordial lithium depends on the present abundances and stellar model modification\cite{1207.3081,1305.6564,Fu,1506.08048}.
Precise measurements have been performed to restrict uncertainties in the thermonuclear rates
for those reactions involved in SBBN\cite{1107.1117,1109.4690,1312.0894,1502.03961}.
SBBN is sensitive to three global parameters:
equivalent species of active neutrinos ($N_{\nu}$), the neutron lifetime ($\tau$), and the number-density ratio of baryons to photons ($\eta$).
A smaller $\eta$ will lower the Li-7 abundance to the observationally allowed range, but in contrast with D/H and He-4.
This dilemma suggests that a simple $\eta$ change cannot explain the Lithium Problem.

Extra non-standard effects during the BBN period --- for example,
extra nuclear processes and the mechanism that destroys Be-7
(The relic Li-7 mainly comes from Be-7 electron-capture decays.)
and produces Li-6 ---
might be possible solutions to the Lithium Problem.
This is related to New Physics (e.g., dark-matter models) beyond the Standard Model of particle physics,
via, for example, new particles or resonances participating in nuclear reactions,
and non-thermal electromagnetic or hadronic energy injection from dark-matter particle decays or annihilations
\cite{1303.0574,1407.0021,1502.01250,1510.08858}.

In previous work\cite{we}, we proposed a possible solution by not straying far from the SBBN model.
We showed that extra contributions from non-thermal particles during and/or shortly after the epoch of BBN,
namely Big Bang nucleosynthesis cosmic rays (BBNCRs), especially hydrogen between 2 and 4 MeV,
can destroy 70\% of the Be-7 via the endothermic reaction $\rBE(p,p\alpha)\rhe$
and successfully allow the Li-7 abundance into the observationally allowed range,
while increasing the Li-6 abundance by only 1 order of magnitude
due to the resonance peak measuring over 2 MeV in the exothermic reaction $\rD(\alpha,\gamma)\rli$,
which is not sufficient.

In this paper, an extension and completion of our preliminary work,
we show whether the BBNCR scenario can phenomenologically account for both Li-7 and Li-6 problems.
In order to increase the Li-6 abundance furthermore, we introduce the endothermic reaction $\rhe(\alpha,p)\rli$,
whose threshold energy is the lowest, 7.048 MeV.
The cross-section of $\rhe(\alpha,p)\rli$ is approximatetly $O(10^6)$ times
than that of the SBBN reaction $\rD(\alpha,\gamma)\rli$\cite{he3a1,he3a2},
so the equivalent energetic helium should be not less than $O(10^{-4})$ times SBBN D,
in order to increase the Li-6 abundance by 2 orders of magnitude.
(In fact, the amount of high-energy He-4 can be lower, $O(10^{-5})$,
because the work-time length of BBNCRs is longer than that of SBBN particles.)
Moreover, energetic helium will also destroy Be-7 via $\rBE(\alpha,p)\rB$, and lower the amount of BBNCRs.
Therefore, we consider not only the hydrogen isotopes as we did previously, but also consider helium isotopes as BBNCRs,
and phenomenologically investigate the combination effect of the parameter space of the amount of BBNCRs and the energy range.

\section{Methods and Modeling}

SBBN assumes that nucleons are in thermal equilibrium with background plasma.
We expect that some kind of plasma turbulence\cite{1605.04646} or other mechanisms
can feed energy to the thermal ions, so nucleons gain energy and deviate from thermal equilibrium.
We proposed a toy model in \cite{we} to show that a kind of electromagnetic acceleration mechanism might be a possibility
(See Appendix A also).

BBNCRs consist of energetic hydrogen, namely protons, deuterons (D), and tritones (T), and helium, namely He-3 and He-4.
Energetic BBNCRs may initiate some endothermic reactions with threshold energy excluded in SBBN, which will change the nuclide abundances.
Therefore, we consider additionally these endothermic reactions in which high-energy hydrogen and helium participate.

The amount of BBNCRs should be low enough to avoid consuming too much D and remain consistent with the SBBN D abundance.
However, the energy of BBNCRs should be high enough to trigger reactions to destroy Be-7 and produce Li-6.

The amount of each kind of BBNCRs must be much lower than that of the corresponding background particles.
Obviously, it may be evolving during their work time with the Universe expanding, but for simplicity and as an average,
we assume that the proportion of accelerated hydrogen to thermal hydrogen is fixed as a single free parameter $\epsilon$
that is unchanging during the BBN period; for helium, see below.

\begin{figure}[htbp]
\centering
\includegraphics[width=.8\textwidth]{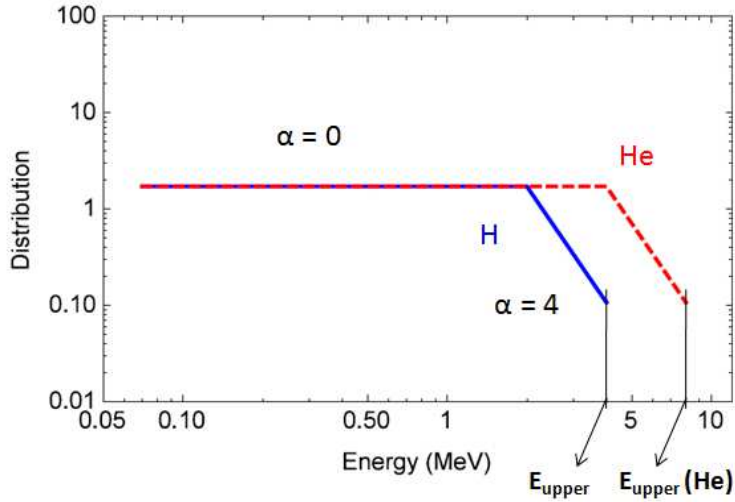}
\flushleft
\caption{BBNCR distribution of energetic hydrogen obeys a broken power law with a cutoff,
with the power index $\alpha=4$, the turning point $E_\text{C}=2$ MeV,
and a sharp upper limit of hydrogen, $E_\text{upper}$.
The energy of helium is 2 times that of hydrogen.
(See Eqs.~(\ref{alpha1}) and (\ref{alpha2}) for the explicit expression.)
\label{spectrum}}
\end{figure}

In the field of cosmic rays, the energy distribution is expected to obey a power law with a ``knee''\cite{knee}:
\begin{equation}
f(E)\propto\frac{E^{-\alpha_1}}{[1+(E/E_\text{C})^{p}]^{(\alpha_2-\alpha_1)/p}}.
\label{knee}
\end{equation}
For simplicity, we assume that the distribution of BBNCR hydrogen obeys a broken power law with a cutoff:
\begin{equation}
f_\text{H}(E)\propto
\begin{cases}
E_\text{C}^0(\text{const.})\ &E<E_\text{C},\\
E^{-\alpha}\ &E_\text{C}<E<E_\text{upper},
\end{cases}
\label{alpha1}
\end{equation}
which is a special case of Eq.~(\ref{knee}) with the power index $\alpha_1=0$, $\alpha_2=\alpha=4$, $p\to\infty$,
and the turning point $E_\text{C}=2$ MeV.
Similar to the GZK cutoff in cosmic rays\cite{GZK1,GZK2},
we introduce a sharp upper limit $E_\text{upper}$, which is a free parameter in our model,
and there is no BBNCR hydrogen once the energy exceeds the upper limit.
The function $f_\text{H}(E)$ is normalized from 2 to 4 MeV:
\begin{equation}
\int_2^4f_\text{H}(E)\,\rd E=1.
\label{normalization}
\end{equation}
Assuming that electric-charged particles are accelerated by electromagnetic force,
and that the gain energy is proportional to the electric charge,
we simply suppose that the energy of helium is 2 times that of hydrogen (See Fig.~\ref{spectrum}.), with the explicit expression
\begin{equation}
f_\text{He}(E)=
\begin{cases}
f_\text{H}(E<E_\text{C})\ &E<2E_\text{C},\\
f_\text{H}(E)\ &2E_\text{C}<E<2E_\text{upper}.
\end{cases}
\label{alpha2}
\end{equation}

Nuclear reactions occur between BBNCRs and background SBBN nuclei in a Boltzmann distribution.
At the same time that BBNCRs destroy Be-7 and produce Li-6, they also initiate other processes,
such as producing Be-7, destroying Li-6, and even destroying D.
The D abundance is a constraint of our model.

Reactions added in our numerical computation are summarized in Table~\ref{total}.
The cross-sections are taken from the experimental data sources DataBase EXFOR\cite{exfor} and ENDF\cite{endf}.
We take the $\rD(p,n)2\rH$ cross-section with a shift of the threshold energy as a substitute for $\rBE(p,p\alpha)\rhe$,
and take the differential cross-section of $\rhe(\alpha,p)\rli$\cite{3HeapLi6} as total cross-section,
for a test.

\begin{table}[htbp]
\caption{Reactions added in our numerical computation.
\label{total}}
\centering
\begin{tabular}{lcl}
\hline
\hline
Process & Threshold & Effect\\
 & (MeV) & \\
\hline
$\rBE(p,p\alpha)\rhe$ & 1.814 & destroys Be-7\\
$\rBE(\alpha,2\alpha)\rhe$ & 2.492 & destroys Be-7\\
$\rBE(\alpha,p)\rB$ & 1.800 & destroys Be-7\\
$\rLI(p,n)\rBE$ & 1.880 & produces Be-7\\
$\rLI(p,\alpha)\ra$ & -- & destroys Li-7\\
$\ra(t,\gamma)\rLI$ & -- & produces Li-7\\
\hline
$\rT(\alpha,n)\rli$ & 8.388 & produces Li-6\\
$\rhe(\alpha,p)\rli$ & 7.048& produces Li-6\\
$\rLI(d,t)\rli$ & 1.278 & produces Li-6\\
$\rBE(d,\rhe)\rli$ & 0.144 & produces Li-6\\
$\rli(\alpha,p)^9$Be & 3.540 & destroys Li-6\\
$\rli(\alpha,d)^8$Be\footnote{$^8$Be decays to 2$\alpha$} & 2.454 & destroys Li-6\\
$\ra(d,\gamma)\rli$ & -- & produces Li-6\\
$\rD(\alpha,\gamma)\rli$ & -- & produces Li-6\\
$\rli(p,\alpha)\rhe$ & -- & destroys Li-6\\
\hline
$\rD(p,n)2\rH$ & 3.337 & destroys D\\
$\rD(\alpha,\alpha)$np & 3.343 & destroys D\\
$\rD(p,\gamma)\rhe$ & -- & destroys D\\
$\rD(d,p)\rT$ & -- & destroys D\\
$\rD(d,n)\rhe$ & -- & destroys D\\
$\rT(d,n)\ra$ & -- & destroys D\\
\hline
\hline
\end{tabular}
\end{table}

\section{Computation and Results}
According to the Boltzmann equation\cite{Iocco}, variation of the abundance of nuclide $i$
through the nuclear reaction,
\begin{equation}
N_ii+N_jj\rightleftharpoons N_kk+N_ll
\end{equation}
(where $N_i$ is the number of particle $i$ that participate in such a reaction),
is described as
\begin{equation}
\frac{\rd Y_i}{\rd t}=\sum_{j,k,l}N_i\left(-\frac{Y_i^{N_i}Y_j^{N_j}}{N_i!N_j!}[ij]_k+\frac{Y_l^{N_l}Y_k^{N_k}}{N_l!N_k!}[lk]_j\right),
\label{Boltzmann}
\end{equation}
where $Y_i\equiv X_i/A_i$ is the abundance of $i$ with $X_i$ the mass fraction, $A_i$ the mass number,
$[ij]$ the rate of destroying $i$, and $[kl]$ the rate of synthesizing $i$.
Element abundances are usually normalized by protons;
for example, $\rD/\rH\equiv Y_{\rD}/Y_{\rH}$.
The sum over $j$ goes through all reactions to destroy $i$,
and the sum over $k,l$ goes through all reactions to synthesize $i$.
The rate $[ij]$ is defined by
\begin{equation}
[ij]\equiv\rho_bN_A\langle ij\rangle=\rho_bN_A\langle\sigma v\rangle,
\label{rate}
\end{equation}
where $\rho_b$ is the baryon energy density,
$N_A$ Avogadro's number,
$\sigma$ the cross-section of the reaction,
and $v$ the relative velocity between the two particles $i$ and $j$.
The $\langle\cdots\rangle$ denotes the average over different relative velocities.
In the case of SBBN, $\langle\cdots\rangle$ denotes the thermal average,
while in the case of BBNCRs it can be computed as
\begin{multline}
\langle\sigma v\rangle(T)=
\frac{1}{K_3}\int_{-1}^1\rd\cos\theta\times\frac{1}{K_1}\int_{-\infty}^{+\infty}f_1(E_1,T)\,\rd E_1\times\\
\epsilon\int^{E_\text{upper}\text{ or }2E_\text{upper}} f_\text{H or He}(E_2)\,\rd E_2\ \sigma(E_i)v(E_1,E_2,\cos\theta),
\label{eq3}
\end{multline}
where $T$ is the Universe's temperature.
The distribution of background particles $f_1(E_1,T)$ is the normalized Maxwellian-Boltzmann distribution,
\begin{equation}
f_1(E_1,T)=2\sqrt{\frac{E_1}{\uppi(kT)^3}}\re^{-E_1/kT},
\end{equation}
where $k$ is Boltzmann's constant
and $E_1$ the energy of the background particle.
Thus, the normalization constant $K_1=1$ and the energy range is $(-\infty,+\infty)$.
Here, $E_2$ is the energy of BBNCRs
and, for the distribution and normalization of BBNCRs, see Eq.~(\ref{normalization}).
When the mass of the background particle is denoted by $m_1$ and that of the BBNCR $m_2$,
we can compute the relative velocity with the angle of incidence $\theta$,
\begin{equation}
v=|\vec{v}_1-\vec{v}_2|=\sqrt{\frac{2E_1}{m_1}+\frac{2E_2}{m_2}-4\sqrt{\frac{E_1E_2}{m_1m_2}}\cos\theta},
\end{equation}
and incident energy $E_i$,
\begin{equation}
E_i=\frac{1}{2}m_iv^2,
\end{equation}
where $m_i$ is the mass of the incident particle.
The normalization constant over $\theta$ is $K_3=\int_{-1}^1\,\rd\cos\theta=2$.

We calculate abundances utilizing the updated version\cite{Kawano1,Kawano2} of the Wagoner code\cite{Wagoner67,Wagoner69} from \cite{code}
with appropriate modification in order to include new contributions from BBNCRs.
We take the $\eta$ value determined from fits to the power spectrum of the Cosmic Microwave Background (CMB)
$\eta=6.19\times10^{-10}$, and the neutron lifetime $\tau=880.1$ sec.
The new contributions are added to the code as new channels.

\begin{figure}[htbp]
\centering
\includegraphics[width=.8\textwidth]{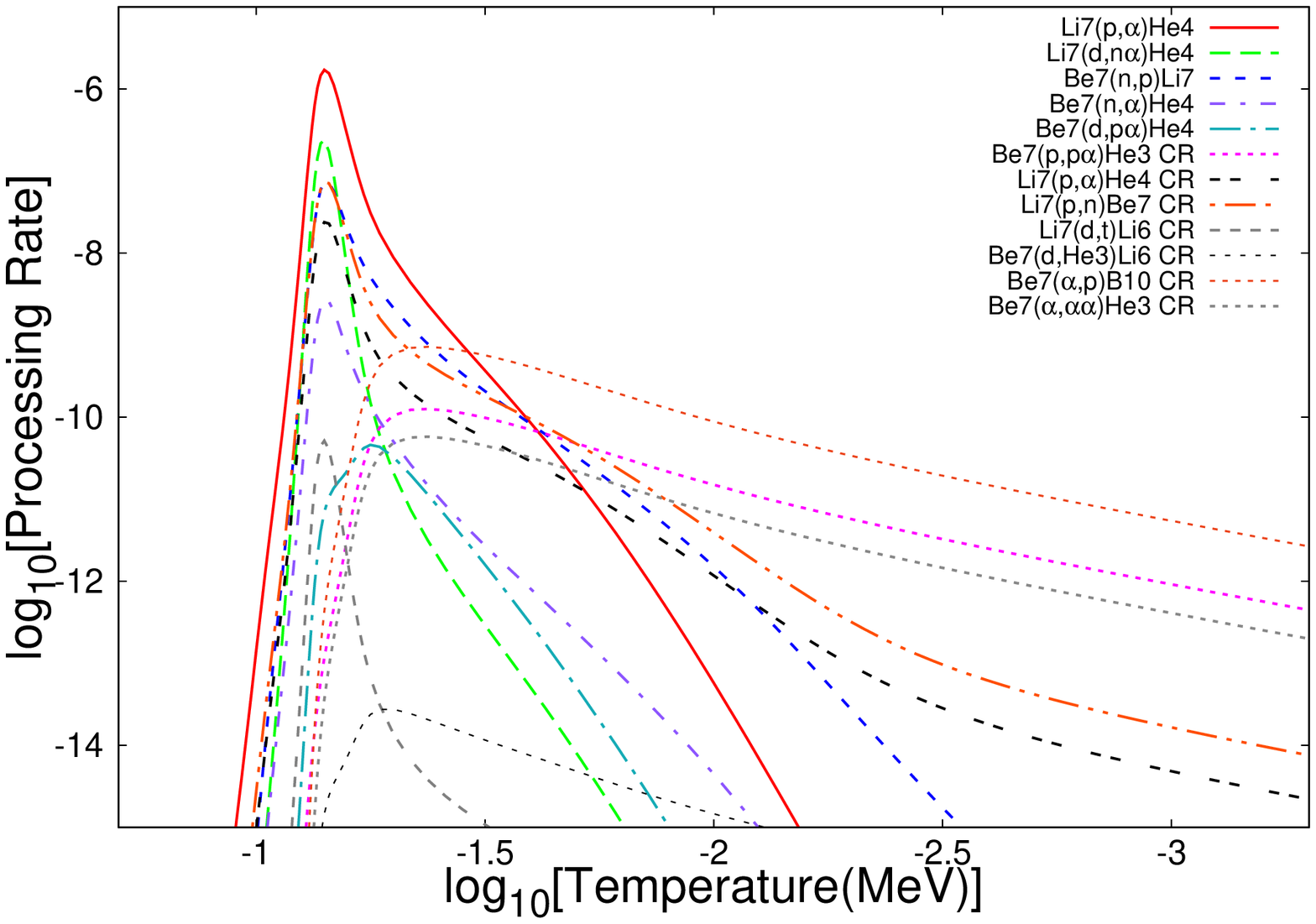}
\flushleft
\caption{Processing rates of the reactions of destroying $\libe$ in the case of BBNCRs
with $\epsilon=1.6\times10^{-5}$ and $E_\text{upper}=3.5$ MeV,
where ``CR'' indicates the reactions that BBNCRs participate in.
\label{Be7desprocessingrate}}
\end{figure}

\begin{figure}[htbp]
\centering
\includegraphics[width=.8\textwidth]{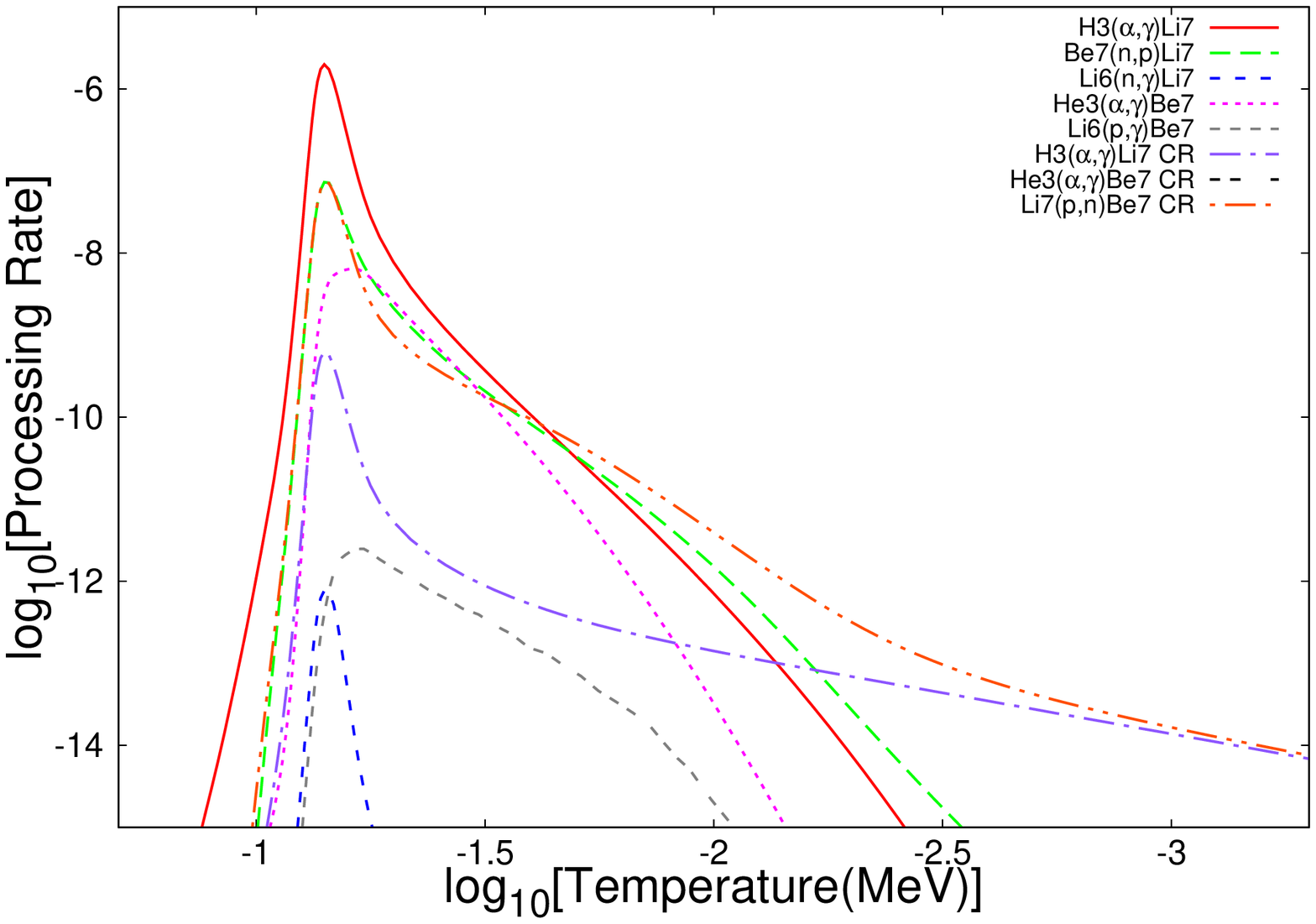}
\flushleft
\caption{Processing rates of the reactions of producing $\libe$ in the case of BBNCRs
with $\epsilon=1.6\times10^{-5}$ and $E_\text{upper}=3.5$ MeV,
where ``CR'' indicates the reactions that BBNCRs participate in.
\label{Be7proprocessingrate}}
\end{figure}

\begin{figure}[htbp]
\centering
\includegraphics[width=.8\textwidth]{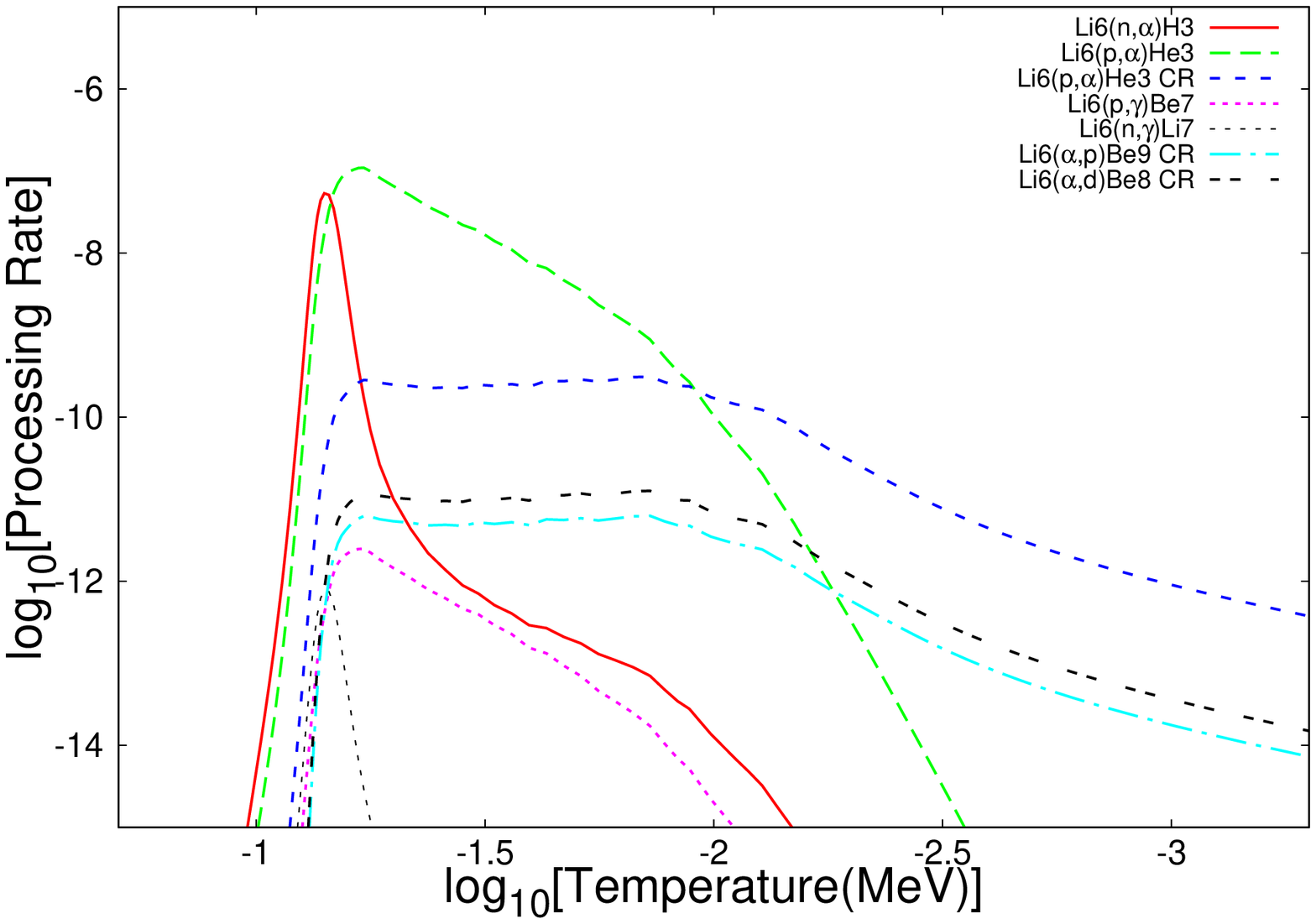}
\flushleft
\caption{Processing rates of the reactions of destroying Li-6 in the case of BBNCRs
with $\epsilon=1.6\times10^{-5}$ and $E_\text{upper}=3.5$ MeV,
where ``CR'' indicates the reactions that BBNCRs participate in.
\label{Li6desprocessingrate}}
\end{figure}

\begin{figure}[htbp]
\centering
\includegraphics[width=.8\textwidth]{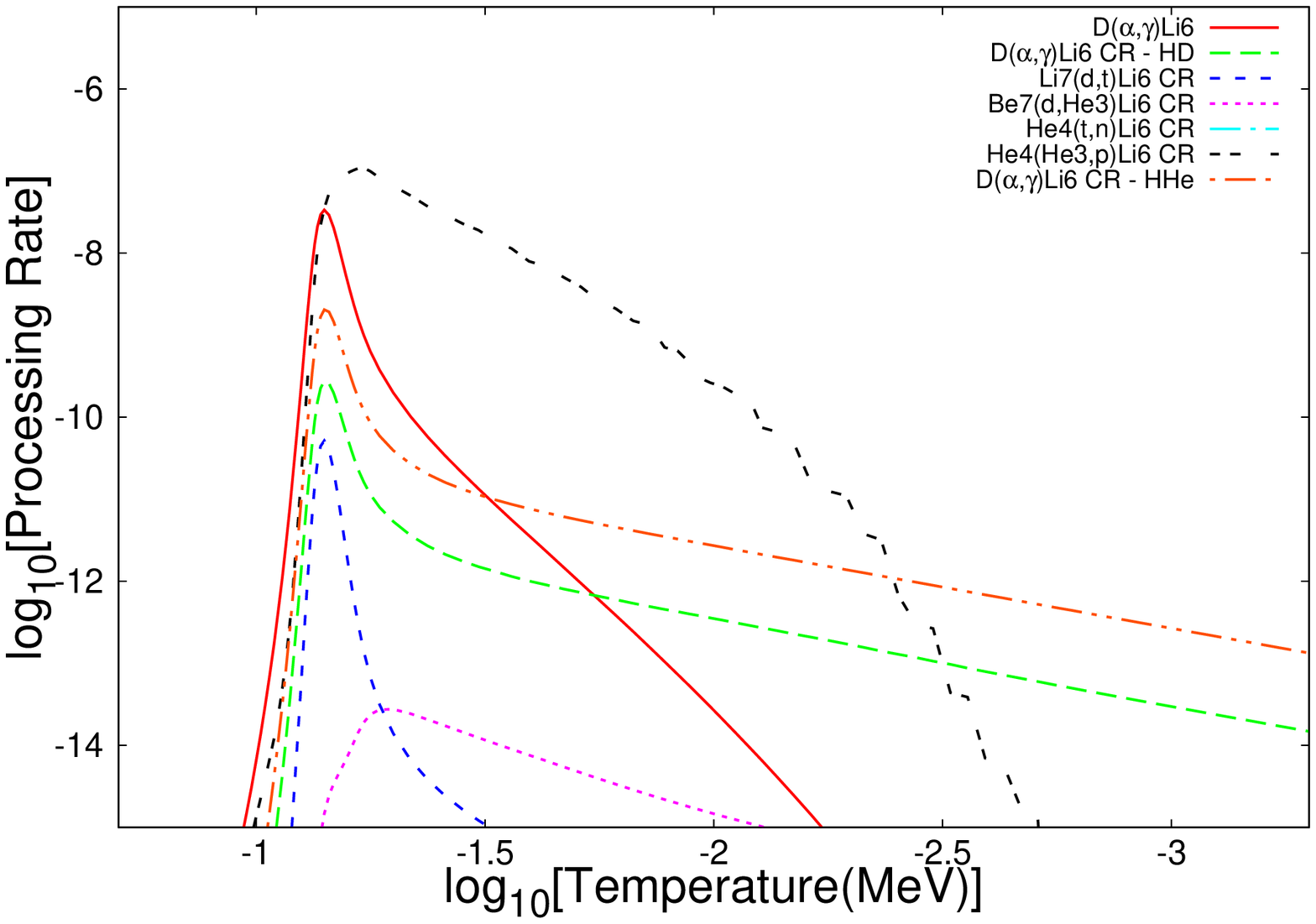}
\flushleft
\caption{Processing rates of the reactions of producing Li-6 in the case of BBNCRs
with $\epsilon=1.6\times10^{-5}$ and $E_\text{upper}=3.5$ MeV,
where ``CR'' indicates the reactions that BBNCRs participate in.
In particular, ``CR-HD'' indicates the reactions in which high-energy D collides with background He-4,
and ``CR-HHe'' indicates the reactions in which high-energy He-4 collides with background D.
\label{Li6proprocessingrate}}
\end{figure}

Processing rates, which are defined as $Y_iY_j[ij]/H$\cite{processingrate}
(See Eq.~(\ref{rate}), and $H$ is the Hubble expansion rate, so $1/H$ indicates the characteristic time length.),
are helpful to see the contribution to the final abundances and the work time of each reaction.
If we choose $\epsilon=1.6\times10^{-5}$ and $E_\text{upper}=3.5$ MeV (See below.),
the processing rates of the destroying and producing $\libe$ reactions in the case of BBNCRs
are shown in Figs.~\ref{Be7desprocessingrate} and \ref{Be7proprocessingrate},
where ``CR'' indicates the reactions that BBNCRs participate in.
It can be seen that:
(1) The contribution of BBNCRs is much lower than
that of SBBN particles when the Universe's temperature is approximately 0.07 MeV.
(2) BBNCRs work later than SBBN reactions, when the Universe's temperature falls below 0.03 MeV.
The processing rates of the reactions of destroying and producing Li-6 in the case of BBNCRs
with $\epsilon=1.6\times10^{-5}$ and $E_\text{upper}=3.5$ MeV
are shown in Figs.~\ref{Li6desprocessingrate} and \ref{Li6proprocessingrate},
where ``CR'' indicates the reactions that BBNCRs participate in.
In particular, ``CR-HD'' indicates the reactions in which high-energy D collides with background He-4,
and ``CR-HHe'' indicates the reactions in which high-energy He-4 collides with background D.
It can be seen that, different from $\libe$,
BBNCR reactions override SBBN ones all the time because of the large cross-section of $\rhe(\alpha,p)\rli$.

\begin{figure}[htbp]
\centering
\includegraphics[width=.8\textwidth]{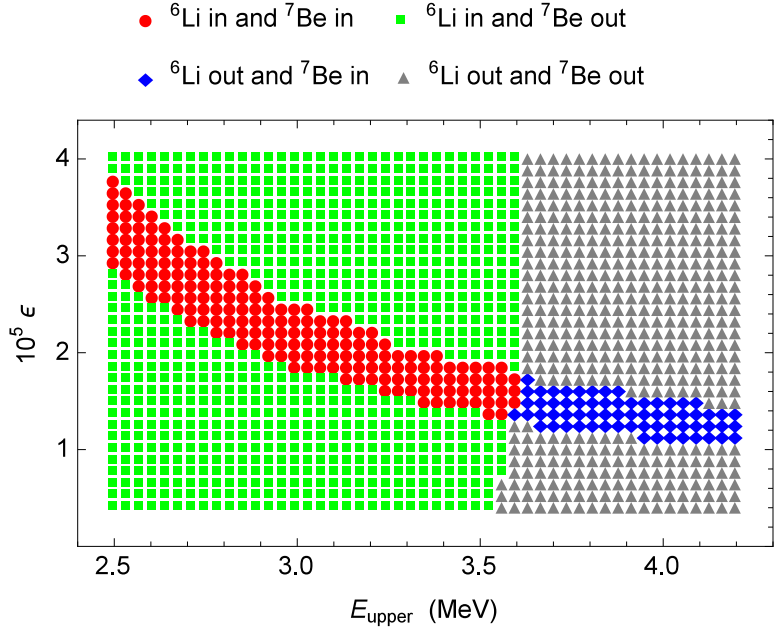}
\flushleft
\caption{Parameter space of $\epsilon$ and $E_\text{upper}$.
We take $\rLI/\rH=(1.6\pm0.3)\times10^{-10}$ and $10^{-13}\leq\rli/\rH\leq0.05\times\rLI/\rH$;
``in'' and ``out'' indicate whether or not the BBNCR predicted abundances fall into the observationally allowed range.
\label{para13}}
\end{figure}

\begin{figure}[htbp]
\centering
\includegraphics[width=.8\textwidth]{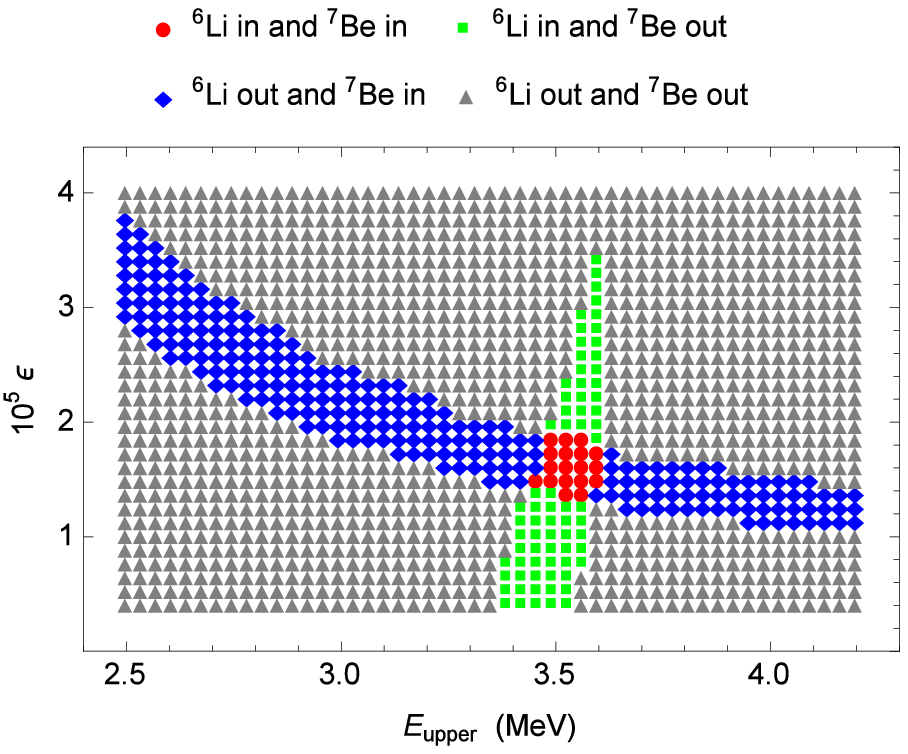}
\flushleft
\caption{Parameter space of $\epsilon$ and $E_\text{upper}$.
We take $\rLI/\rH=(1.6\pm0.3)\times10^{-10}$ and $10^{-12}\leq\rli/\rH\leq0.05\times\rLI/\rH$;
``in'' and ``out'' indicate whether or not the BBNCR predicted abundances fall into the observationally allowed range.
\label{para12}}
\end{figure}

There are two parameters in our model: $\epsilon$ and $E_\text{upper}$.
The former represents the amount of BBNCRs, and the latter indicates the highest energy that high-energy hydrogen achieves.
According to observations,
we scan the two parameters to find the observationally allowed space;
that is, we take the Li-7 abundance $\rLI/\rH=(1.6\pm0.3)\times10^{-10}$
and the upper limit of the Li-6 abundance $\rli/\rLI\leq0.05$\cite{1003.4510} given in Particle Data Group (2014)\cite{pdg}.
Without a clear lower limit of the Li-6 abundance, we show the results in two cases for reference,
$10^{-13}\leq\rli/\rH\leq0.05\times\rLI/\rH$ and $10^{-12}\leq\rli/\rH\leq0.05\times\rLI/\rH$, respectively,
in Figs.~\ref{para13} and \ref{para12}.
In the figures, ``in'' and ``out'' indicate whether or not the BBNCR predicted abundances fall into the observationally allowed range.

\begin{figure}[htbp]
\centering
\includegraphics[width=.8\textwidth]{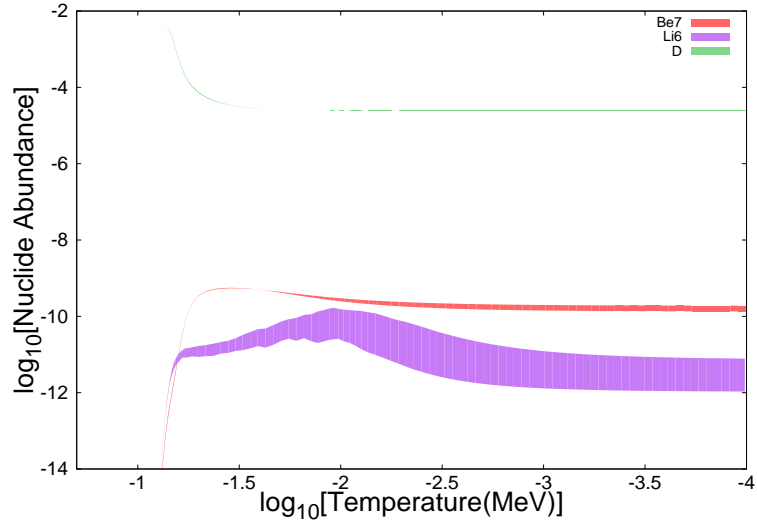}
\flushleft
\caption{D, Be-7, and Li-6 abundances as a function of the Universe's temperature
with the observationally allowed $\epsilon$ and $E_\text{upper}$ ranges;
namely, those shown in Fig.~\ref{para12}.
\label{abundbeli}}
\end{figure}

It can be seen that there is a parameter space with which to reconcile the contradictory aspects of Li-7 and Li-6;
meanwhile, the parameter space is almost harmless with respect to the D abundance.
Percentages of $\epsilon$ deviation and Li-7 destroyed are of the same order of magnitude.
For example, a 10\% increase of $\epsilon$ leads to 10\% more Li-7 destroyed.
For Li-7, $\epsilon$ and $E_\text{upper}$ are substitutions:
higher $E_\text{upper}$ means less $\epsilon$.
However for Li-6, energetic helium promotes Li-6 significantly.
Once the reaction $\rhe(\alpha,p)\rli$ is triggered by BBNCR helium, Li-6 can be complemented,
so the Li-6 abundance is more sensitive to $E_\text{upper}$ than to $\epsilon$.
The evolution of D, Be-7, and Li-6 abundances as a function of the Universe's temperature
with the two BBNCR parameters allowed by Li-7 and Li-6 abundance observations,
namely those shown in Fig.~\ref{para12},
is shown in Fig.~\ref{abundbeli}.
It can be seen that the D abundance changes little.

\begin{figure}[htbp]
\centering
\includegraphics[width=.8\textwidth]{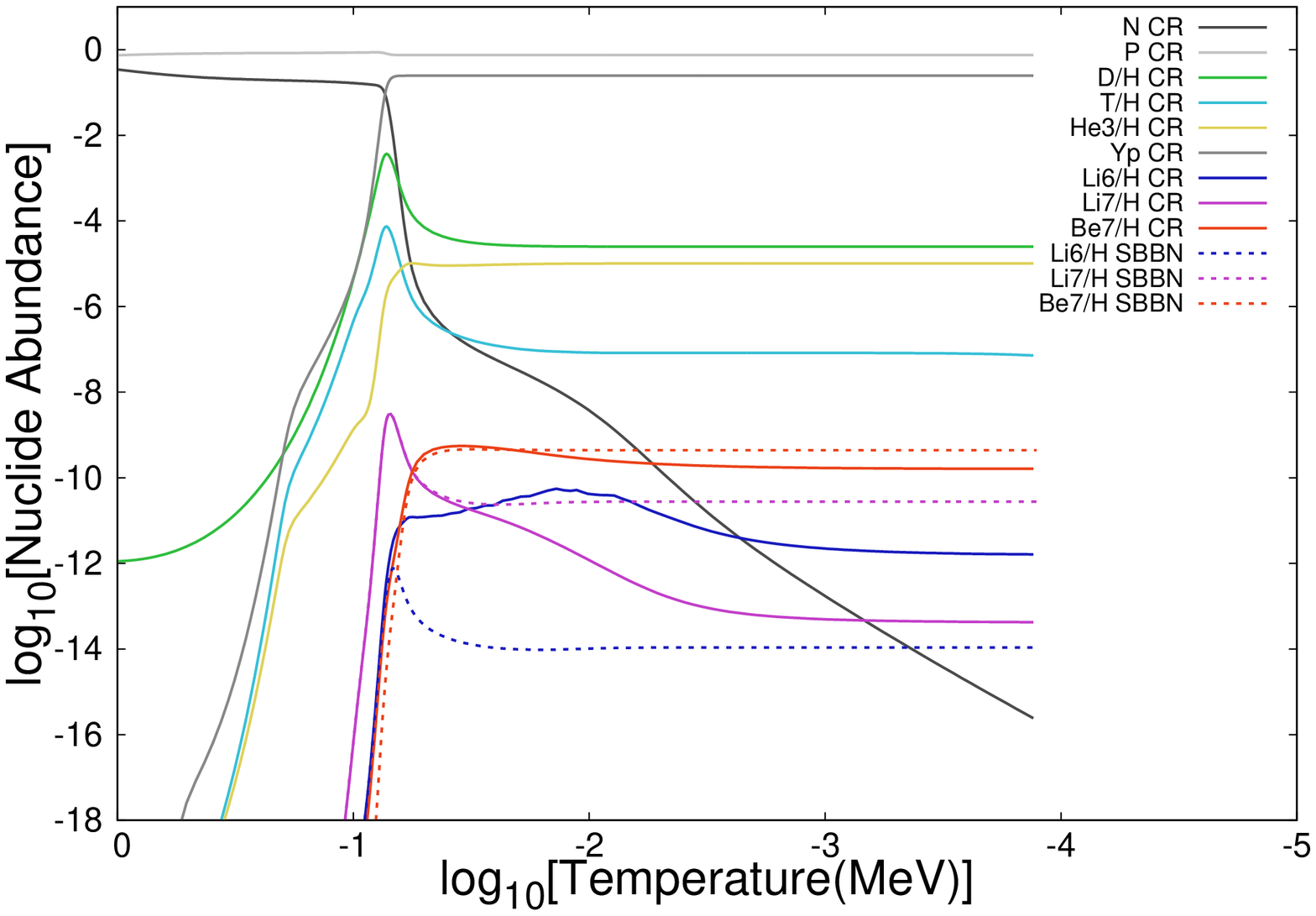}
\flushleft
\caption{Element abundances as a function of the Universe's temperature
with $\epsilon=1.6\times10^{-5}$ and $E_\text{upper}=3.5$ MeV.
Solid lines denote the BBNCR results and dashed lines those for SBBN.
\label{evolution}}
\end{figure}

The evolution of all light element abundances as a function of the Universe's temperature is shown in Fig.~\ref{evolution}
with, for example, $\epsilon=1.6\times10^{-5}$ and $E_\text{upper}=3.5$ MeV.
The solid lines denote the BBNCR results and the dashed lines those for SBBN.

Many groups also give lithium abundance observation results,
and the maximum and minimum of allowed parameters
according to other choices of lithium abundance observations\cite{Spite,9905211,0510636,0610245,1005.2944}
are listed in Table~\ref{diffpara}.
It is not difficult to see that there is an allowed parameter space for each choice.
These allowed ranges do not coincide because there is no overlap of all choices of lithium abundance observations.

The cross-sections of $\rBE(p,p\alpha)\rhe$ and $\rhe(\alpha,p)\rli$ are uncertain.
We test the effect of cross-section uncertainty in two cases:
(1) The trial cross-sections time 1/10.
(2) The cross-sections are constant, using the value once the energy exceeds the threshold.
(The former is 0.22 mb and the latter is 5.2 mb.)
The maximum and minimum of the allowed parameters are listed in Table~\ref{diffpara}.
It can be seen that the cross-section uncertainty will change the parameters only little, relative to the cross-section changes;
that is, our BBNCR hypothesis is insensitive to the amplitude or resonance peak of the uncertain cross-sections.

Another test is on the energy spectrum.
The explicit energy spectrum shape is fixed, and we test some other shape choices.
We suppose that the energy distribution of BBNCRs ($f_\text{He}(E)=f_\text{H}(E)$) is uniform all through;
namely the power index $\alpha=0$ even for $E_\text{C}<E<E_\text{upper}$,
and is still normalized from 2 to 4 MeV.
Moreover, the upper limit of helium energy is 2 times that of hydrogen, as before.
The maximum and minimum of the allowed parameters are listed in Table~\ref{diffpara}.
It can be seen that our hypothesis is also insensitive to the explicit shape of the BBNCR energy spectrum.

\begin{table}[htbp]
\caption{Maximum and minimum of allowed parameters
according to other choices of lithium abundance observations, cross-sections, or energy spectrum.
\label{diffpara}}
\centering
\begin{tabular}{lcc}
\hline
\hline
Choice & $\epsilon$ range & $E_\text{upper}$ range\\
& ($10^{-5}$) & (MeV)\\
\hline
\multicolumn{3}{c}{a. Other choices of lithium abundance observations.}\\
\hline
$\rLI/\rH=(1.6\pm0.3)\times10^{-10}$ and & & \\
$10^{-12}\leq\rli/\rH\leq0.05\times\rLI/\rH$\cite{1003.4510}. & $[1.36,1.84]$ & $[3.46,3.60]$\\
\hline
$\rLI/\rH=(1.23^{+0.68}_{-0.32})\times10^{-10}$ and & & \\
$10^{-12}\leq\rli/\rH\leq0.05\times\rLI/\rH$\cite{9905211}. & $[1.36,2.32]$ & $[3.46,3.60]$\\
\hline
$\rLI/\rH=(1.26\pm0.26)\times10^{-10}$ and & & \\
$10^{-12}\leq\rli/\rH\leq0.05\times\rLI/\rH$\cite{0610245}. & $[1.60,2.20]$ & $[3.49,3.60]$\\
\hline
$\rLI/\rH=(1.86\pm0.23)\times10^{-10}$ and & & \\
$10^{-12}\leq\rli/\rH\leq0.05\times\rLI/\rH$\cite{1005.2944}. & $[1.24,1.48]$ & $[3.46,3.56]$\\
\hline
\hline
\multicolumn{3}{c}{b. Test on cross-sections}\\
\hline
The $\rBE(p,p\alpha)\rhe$ and $\rhe(\alpha,p)\rli$ & &\\
trial cross-sections time 1/10. & &\\
$\rLI/\rH=(1.6\pm0.3)\times10^{-10}$ and & & \\
$10^{-12}\leq\rli/\rH\leq0.05\times\rLI/\rH$. & $[1.84,2.20]$ & $[3.73,3.82]$\\
\hline
The $\rBE(p,p\alpha)\rhe$ cross-section & &\\
is 0.22 mb and the $\rhe(\alpha,p)\rli$ & &\\
cross-section is 5.2 mb, once & &\\
the energy exceeds the threshold. & &\\
$\rLI/\rH=(1.6\pm0.3)\times10^{-10}$ and & & \\
$10^{-12}\leq\rli/\rH\leq0.05\times\rLI/\rH$. & $[1.90,2.20]$ & $[3.52,3.65]$\\
\hline
\hline
\multicolumn{3}{c}{c. Test on energy spectrum}\\
\hline
$f_\text{He}(E)=f_\text{H}(E)$ is uniform all through. & &\\
$\rLI/\rH=(1.6\pm0.3)\times10^{-10}$ and & & \\
$10^{-12}\leq\rli/\rH\leq0.05\times\rLI/\rH$. & $[1.24,1.72]$ & $[3.28,3.39]$\\
\hline
\hline
\end{tabular}
\end{table}

\section{Conclusions and Discussion}
As an extension and completion of our preliminary work,
a BBNCR solution that does not stray far from the SBBN model and is different from the New Physics sectors
is proposed in this paper.
In the case in which BBNCRs consist of both hydrogen and helium isotopes,
we find that there an allowed parameter space of the amount of BBNCRs ($\epsilon$)
and the highest energy that high-energy hydrogen achieves ($E_\text{upper}$) exists
to explain the discrepancy between theoretically predicted Li-7 and Li-6 abundances and astronomical observations,
while other element abundances change little.
A recommended parameter choice is $\epsilon=1.6\times10^{-5}$ and $E_\text{upper}=3.5$ MeV.
In addition, the $\rBE(p,p\alpha)\rhe$ and $\rhe(\alpha,p)\rli$ cross-section uncertainty will also change parameters little,
relative to the cross-section changes.
It is worth mentioning that the explicit energy spectrum in Fig.~\ref{spectrum} may be not necessary,
since our hypothesis is insensitive to the explicit shape of the BBNCR energy spectrum.
However, the highest energy that BBNCRs achieve and the amount of BBNCRs matters.
The amount of BBNCRs and of Li-7 destroyed are nearly proportional:
for example, a 10\% increase of BBNCRs leads to 10\% more Li-7 destroyed.
Once the reaction $\rhe(\alpha,p)\rli$ is triggered by BBNCRs with enough energy, Li-6 can be complemented.

Under our hypothesis, the D abundance changes little (about 0.4\% destroyed),
which is safer than the situation in which BBNCRs consist of only hydrogen discussed in our previous work\cite{we},
because of the lower the amount of BBNCRs when high-energy helium is considered.
BBNCRs may promote B-10 via $\rBE(\alpha,p)\rB$ to $10^{-12}$,
approaching the upper limit of observations\cite{boron},
and may also produce more Be-9, C-12, and C-13 than SBBN.
A larger amount of primordial CNO would affect the first generation of stars
(or Population \uppercase\expandafter{\romannumeral3})\cite{1403.6694}.
Precise observations may or may not support our hypothesis.

The BBNCR source is still an open question.
For example, just as with a $10^{-5}$ turbulence in CMB,
it seems natural that there is a $10^{-5}$ turbulence of plasma flux (namely $O(10^{-4})$ of total energy) during BBN,
regardless of acceleration or energy injection mechanism.
Clarifying the acceleration or energy injection mechanism is beyond the scope of this paper and deserves further study;
specific goals of future work include:
(1) How the required amount can be induced;
(2) How the required energy can be achieved;
(3) How the BBNCRs vary during their work time.

\section*{Acknowledgements}
This work was supported in part by the Natural Science Foundation of China (Grants. Nos. 11635001, 11375014, and 11475121).

\appendix
\section{A toy model for the source of BBNCRs}
Here we propose a toy model to show that
a kind of electromagnetic acceleration mechanism might be a possibility
for the thermal nucleons to gain energy and deviate from thermal equilibrium.
(Also see \cite{we}.)
The primordial magnetic fields might be created at some early stage of the evolution of the Universe,
e.g., inflation, the electro-weak phase transition, quark-hadron phase transition and so on.
As investigated in \cite{9602031}, after electro-weak phase transition
the magnetic field builds up and evolves with the expanding Universe.
We can estimate the strength of induced electric field through $E\approx\Delta B/\Delta t\times L\approx B H L$,
where $B\propto R^{-2}\propto T^2$ is the characteristic magnetic field,
$H$ the Hubble expansion rate,
and $R$ the cosmic scale factor.
Here $L=l_0(R/R_0)\propto T^{-1}$ is the characteristic length of the magnetic field turbulence, in which the factor $R/R_0$ indicates the effect of cosmic expansion.
A little later than the electro-weak phase transition ($T_{\text{EW}}=100$ GeV),
$B_{\text{EW}}=10^{14}$ T, $L_{\text{EW}}=10^{-5}\times10^{4}=10^{-1}$ m,
using such initial parameters and extrapolating to, for example, 0.03 MeV,
we get the induced electric field $E\approx10^3$ V/m when the Universe's temperature $T=0.03$ MeV.
So an electric-charged particle such as hydrogen and helium
which is undergoing such electric field
will gain energy $\Delta E\approx qEL_{\text{free}}=qEv_\varparallel\tau_{\text{ther}}$,
where $q$ is the electric charge of the particle,
$L_{\text{free}}$ the free path of an energetic (O(MeV)) particle,
$v_\varparallel$ the particle velocity parallel to the electric field,
$\tau_{\text{ther}}$ the thermalization time.
Since $v_\varparallel=10^{-2}c$ ($c$ is the light velocity), $\tau_{\text{ther}}\approx10^{-4}$ sec\cite{magnetic},
the possible energy gain of an accelerated proton $\Delta E\approx10^5$ eV,
and that of a helium is 2 times.
Charged particles have a opportunity to be accelerated several times and thus to gain O(MeV) energy.
Therefore, the energy spectrum will not be too hard.

\section{Summery of ignored nuclear reactions}
Exothermic reactions, and endothermic reactions,
with threshold energy, for example, below 5 MeV that energetic hydrogen (p, D, and T) participates in,
and with threshold energy, then, below 10 MeV that helium (He-3 and He-4) participates in, are summerized here.
Since there is no stable element with mass number $A=5$ or $A=8$, the following element decays are worth mentioning:
(1) T decays to He-3,
and considering no constraint on the primordial He-3 abundance,
reactions that destroy and produce T/He-3 simply are not added in our numerical computation;
(2) He-6 decays to Li-6;
(3) Li-8 decays to 2$\alpha$;
(4) B-8 decays to 2$\alpha$;
(5) Li-9 decays to 2$\alpha$ or to Be-9;
(6) Be-10 decays to B-10;
(7) C-10 decays to B-10;
(8) C-11 decays to B-11.

\subsection{Ignored exothermic reactions}
The following exothermic reactions that energetic hydrogen and helium participate in may produce elements with mass number $A\ge9$,
such as Be-9, B-10, B-11:
$\rli+\rT\rightarrow\gamma+{}^9$Be,
$\rLI+\rD\rightarrow\gamma+{}^9$Be,
$\rLI+\rT\rightarrow\gamma+\rB$,
$\rLI+\rT\rightarrow\rn+{}^9$Be,
$\rBE+\rT\rightarrow\gamma+\rB$,
$\rBE+\rT\rightarrow\rrp+{}^9$Be,
$\rLI+\rhe\rightarrow\gamma+\rB$,
$\rLI+\rhe\rightarrow\rrp+{}^9$Be,
$\rBE+\rhe\rightarrow\gamma+{}^{10}$C,
$\rBE+\ra\rightarrow\gamma+{}^{11}$C.

Besides those in Table~\ref{total} and above,
the following exothermic reactions that energetic hydrogen and helium participate in are less sufficient,
so they are ignored and not added in our numerical computation:

$\rrp+\rn\rightarrow\gamma+\rD$,
$\rD+\rn\rightarrow\gamma+\rT$,
$\rhe+\rn\rightarrow\rrp+\rT$,
$\rT+\rrp\rightarrow\gamma+\ra$,
$\rli+\rrp\rightarrow\gamma+\rBE$,
$\rLI+\rD\rightarrow\rn+2\ra$,
$\rBE+\rD\rightarrow\rrp+2\ra$,
$\rBE+\rrp\rightarrow\gamma+{}^8$B,
$\rD+\rD\rightarrow\gamma+\ra$,
$\rT+\rT\rightarrow2\rn+\ra$,
$\rT+\rT\rightarrow\gamma+{}^6$He,
$\rhe+\rT\rightarrow\gamma+\rli$,
$\rhe+\rT\rightarrow\rD+\ra$,
$\rhe+\rT\rightarrow\rn+\rrp+\ra$,
$\rli+\rD\rightarrow2\ra$,
$\rli+\rD\rightarrow\rrp+\rLI$,
$\rli+\rD\rightarrow\rn+\rBE$,
$\rli+\rD\rightarrow\rrp+\rT+\ra$,
$\rli+\rD\rightarrow\rn+\rhe+\ra$,
$\rli+\rT\rightarrow\rn+2\ra$,
$\rli+\rT\rightarrow\rD+\rLI$,
$\rli+\rT\rightarrow\rrp+{}^8$Li,
$\rLI+\rT\rightarrow\ra+{}^6$He,
$\rLI+\rT\rightarrow2\rn+2\ra$,
$\rBE+\rT\rightarrow\ra+\rli$,
$\rBE+\rT\rightarrow\rD+2\ra$,
$\rBE+\rT\rightarrow\rrp+\rn+2\ra$,
$\rBE+\rT\rightarrow\rhe+\rLI$;

$\rD+\rhe\rightarrow\rrp+\ra$,
$\rT+\rhe\rightarrow\gamma+\rli$,
$\rT+\rhe\rightarrow\rD+\ra$,
$\rT+\rhe\rightarrow\rn+\rrp+\ra$,
$\rhe+\rhe\rightarrow2\rrp+\ra$,
$\rli+\rhe\rightarrow\rrp+2\ra$,
$\rli+\rhe\rightarrow\rD+\rBE$,
$\rLI+\rhe\rightarrow\ra+\rli$,
$\rLI+\rhe\rightarrow\rD+2\ra$,
$\rLI+\rhe\rightarrow\rn+\rrp+2\ra$,
$\rBE+\rhe\rightarrow2\rrp+2\ra$,
$\rhe+\ra\rightarrow\gamma+\rBE$.

\subsection{Ignored endothermic reactions}
The following endothermic reactions with threshold energy below 5 MeV that energetic hydrogen participates in,
and with threshold energy below 10 MeV that energetic helium participates in,
may produce elements with mass number $A\ge9$,
such as Be-9, B-10:
$\rli(\alpha,p)^9$Be ($E_\text{th}=3.5$ MeV),
$\rLI+\ra\rightarrow\rrp+{}^{10}$Be ($E_\text{th}=4.0$ MeV),
$\rLI+\ra\rightarrow\rn+\rB$ ($E_\text{th}=8.8$ MeV),
$\rBE+\ra\rightarrow\rn+{}^{10}$C ($E_\text{th}=8.8$ MeV).

Besides those in Table~\ref{total} and above,
the following endothermic reactions that energetic hydrogen and helium participate in are less sufficient,
so they are ignored and not added in our numerical computation:

$\rT+\rrp\rightarrow\rn+\rhe$ ($E_\text{th}=1.0$ MeV),
$\rli+\rrp\rightarrow\rrp+\rD+\ra$ ($E_\text{th}=1.7$ MeV),
$\rli+\rrp\rightarrow\rn+2\rrp+\ra$ ($E_\text{th}=4.3$ MeV),
$\rLI+\rrp\rightarrow\rrp+\rT+\ra$ ($E_\text{th}=2.8$ MeV),
$\rLI+\rrp\rightarrow\rn+\rhe+\ra$ ($E_\text{th}=3.7$ MeV),
$\rT+\rD\rightarrow2\rn+\rhe$ ($E_\text{th}=5.0$ MeV),
$\rhe+\rD\rightarrow2\rrp+\rT$ ($E_\text{th}=2.4$ MeV),
$\rli+\rD\rightarrow2\rD+\ra$ ($E_\text{th}=2.0$ MeV),
$\rli+\rD\rightarrow\rn+\rrp+\rD+\ra$ ($E_\text{th}=4.9$ MeV),
$\rLI+\rD\rightarrow\rD+\rT+\ra$ ($E_\text{th}=3.2$ MeV),
$\rLI+\rD\rightarrow2\rn+\rBE$ ($E_\text{th}=5.0$ MeV),
$\rBE+\rD\rightarrow2\rrp+\rLI$ ($E_\text{th}=0.7$ MeV),
$\rBE+\rD\rightarrow\rD+\rhe+\ra$ ($E_\text{th}=2.0$ MeV),
$\rBE+\rD\rightarrow2\rn+{}^8$B ($E_\text{th}=2.7$ MeV),
$\rBE+\rD\rightarrow2\rrp+\rT+2\ra$ ($E_\text{th}=3.9$ MeV),
$\rBE+\rD\rightarrow\rn+\rrp+\rhe+\ra$ ($E_\text{th}=4.9$ MeV),
$\rli+\rT\rightarrow\rn+\rrp+\rLI$ ($E_\text{th}=1.8$ MeV),
$\rli+\rT\rightarrow\rD+\rT+\ra$ ($E_\text{th}=2.2$ MeV),
$\rli+\rT\rightarrow2\rn+\rBE$ ($E_\text{th}=4.3$ MeV),
$\rLI+\rT\rightarrow2\rT+\ra$ ($E_\text{th}=3.5$ MeV),
$\rBE+\rT\rightarrow\rT+\rhe+\ra$ ($E_\text{th}=2.3$ MeV);

$\rD+\rhe\rightarrow2\rrp+\rT$ ($E_\text{th}=3.6$ MeV),
$\rD+\rhe\rightarrow\rn+\rrp+\rhe$ ($E_\text{th}=5.6$ MeV),
$\rli+\rhe\rightarrow2\rrp+\rLI$ ($E_\text{th}=0.7$ MeV),
$\rli+\rhe\rightarrow\rD+\rhe+\ra$ ($E_\text{th}=2.2$ MeV),
$\rli+\rhe\rightarrow\rn+{}^8$B ($E_\text{th}=3.0$ MeV),
$\rli+\rhe\rightarrow\rn+\rrp+\rBE$ ($E_\text{th}=3.2$ MeV),
$\rli+\rhe\rightarrow2\rrp+\rT+\ra$ ($E_\text{th}=4.4$ MeV),
$\rli+\rhe\rightarrow\rn+\rrp+\rhe+\ra$ ($E_\text{th}=5.5$ MeV),
$\rLI+\rhe\rightarrow\rT+\rBE$ ($E_\text{th}=1.3$ MeV),
$\rLI+\rhe\rightarrow\rT+\rhe+\ra$ ($E_\text{th}=3.5$ MeV),
$\rLI+\rhe\rightarrow2\rrp+{}^8$Li ($E_\text{th}=8.1$ MeV),
$\rLI+\rhe\rightarrow\rrp+\rT+\rli$ ($E_\text{th}=9.3$ MeV),
$\rBE+\rhe\rightarrow2\rhe+\ra$ ($E_\text{th}=2.3$ MeV),
$\rBE+\rhe\rightarrow\rD+{}^8$B ($E_\text{th}=7.7$ MeV),
$\rBE+\rhe\rightarrow\rrp+\rhe+\rli$ ($E_\text{th}=8.0$ MeV),
$\rBE+\rhe\rightarrow3\rrp+\rLI$ ($E_\text{th}=8.7$ MeV),
$\rD+\ra\rightarrow\rn+\rrp+\ra$ ($E_\text{th}=6.6$ MeV),
$\rli+\ra\rightarrow\rn+\rrp+2\ra$ ($E_\text{th}=6.2$ MeV),
$\rLI+\ra\rightarrow\rT+2\ra$ ($E_\text{th}=3.9$ MeV),
$\rBE+\ra\rightarrow\rrp+\ra+\rli$ ($E_\text{th}=8.8$ MeV).

\bibliography{paraspace}

\end{document}